\documentclass{ap-jnmp}

\usepackage{amsfonts, epsfig}

\usepackage{color}

\newcommand{\bea}{\begin{eqnarray}}
\newcommand{\eea}{\end{eqnarray}}
\newcommand{\beq}{\begin{equation}}
\newcommand{\eeq}{\end{equation}}
\newcommand{\enn}{\nonumber \end{equation}}

\copyrightauthor{}

\title{Generalized Virasoro algebra: left-symmetry and related algebraic and hydrodynamic 
properties}

\author{Mahouton Norbert Hounkonnou$^\dag${\footnote{Corresponding author: with copy to hounkonnou@yahoo.fr}}, \\
 Partha Guha $^\ddag$
  \\
 and  Tudor Ratiu$^a$}

\address{$^\dag$University of Abomey-Calavi, International Chair in Mathematical Physics
and Applications (ICMPA-UNESCO Chair),  072 B.P. 50 Cotonou, Republic of Benin\\
$^\ddag$S. N. Bose National Centre for Basic Sciences, JD Block, Sector - 3, Salt Lake, Calcutta - 700098, India\\
$^a$Section de Math\'ematiques, Ecole Polytechnique F\'ed\'erale de Lausanne, 1015 Lausanne, Switzerland\\
\email{$^\dag$norbert.hounkonnou@cipma.uac.bj, \\
 $^\ddag$parthaguh@gmail.com,\\
 $^a$tudor.ratiu@epfl.ch}}

\begin{document}

\maketitle
\thispagestyle{empty}

\today

\vphantom{\vbox{%
\begin{history}
\received{(Day Month Year)}
\revised{(Day Month Year)}
\accepted{(Day Month Year)}
%\comby{(xxxxxxxxx)}
\end{history}
}}

\begin{abstract}
Motivated by the work of Kupershmidt (J. Nonlin. Math. Phys.  {\bf 6} (1998), 222 --245) 
we discuss the occurrence of  left symmetry  in a generalized Virasoro algebra.
%{
The multiplication rule is defined, which is necessary and sufficient  for this algebra to be
  quasi-associative.
%}.
 Its link to geometry and nonlinear systems of hydrodynamic type is also recalled.
Further, the criteria of skew-symmetry, derivation and Jacobi identity making this algebra 
into  a Lie  algebra  are derived. The coboundary operators are defined and discussed. We deduce the hereditary operator and its generalization to the corresponding $3-$ary bracket. Further, we derive the so-called $\rho-$compatibility equation and perform a  phase-space extension. 
Finally, concrete relevant particular cases are investigated.
\end{abstract}

\ccode{Mathematics Subject Classification (MSc) 2010: 17B68; 17B66}
%\today
  
%\tableofcontents

\section{Introduction}

The Virasoro algebra, also known as centrally extended Witt algebra, is probably one of the most important
algebra studied by physicists and mathematicians in last few decades. It has a profound 
impact on  mathematical and physical sciences. It appears naturally in problem with conformal symmetry and
where the essential space-time is one or two dimensional and space is compactified to a circle. For more details, see \cite{auslander}--\cite{burde}, \cite{francesco}, \cite{fuchs}, \cite{goddard}, \cite{witten}, \cite{raina}, \cite{bai}--\cite{palcoux}, \cite{virasoro}, \cite{wass} but also references therein.

\noindent
We deal with one of the most important infinite dimensional Lie algebra, the Witt algebra ${\mathcal W}$ and its
universal central extension.
The Witt algebra is defined as the complex Lie algebra of derivations of the algebra ${\Bbb C}[\theta, 
\theta^{-1}]$ of complex Laurent polynomials. The elements of Witt algebra ${\mathcal W}$ are defined as
$d_n= ie^{i n\theta}{d\over {d\theta}},\;\;n\in \mathbb Z$, so  
$$ {\mathcal W} = \oplus_{n \in \mathbb Z}{\mathbb C} d_n. $$
The Lie-bracket of elements of ${\mathcal W}$ yields $[d_m, d_n]= (m-n)d_{m+n}.$ 

The Virasoro algebra is constructed from the Witt algebra ${\mathcal W}$ by non-trivial central extension,
called the Gelfand-Fuchs cocycle. 
%Thus 
%we add a center $c$ to the basis of the Witt algebra, which commutes with everything. The Virasoro algebra has a basis
%$\{c\} \cup \{d_m | m \in {\Bbb Z} \}$, such that
%$$
%[c,d_m ] = 0, \qquad [d_m,d_n] = (m-n)d_{m+n} + \delta_{m,-n}\frac{m^3 -m}{12}c.
%$$

\bigskip

{

Recently, Kuperschmidt \cite{kuper1} investigated the Virasoro algebra with the multiplication

\bea\label{virkuper}
[e_p, e_q]&:=&e_p\star e_q -  e_q\star e_p
=(p - q) e_{p+q} + \theta (p^3 - p) \delta_{p+q},\;\; p, q\in \mathbb Z,\cr
[\theta, e_p]&=&0,
\eea
in a quasiassociative algebra endowed with the product

\bea
e_p\star e_q&=& -{{q(1+\epsilon q)}\over{1+\epsilon(p+q)}}e_{p+q}+{1\over 2}\theta[p^3 - p +(\epsilon - \epsilon^{-1})p^2]\delta_{p+q}^0,\cr
e_p\star\theta&=& \theta\star e_p=0.
\eea

He focussed his analysis on  the centerless quasiassociative multiplication

\bea
e_p\star e_q&=& -{{q(1+\epsilon q)}\over{1+\epsilon(p+q)}}e_{p+q}.
\eea
He  verified that this multiplication satisfies the quasiassociativity property
\bea
e_p\star( e_q\star  e_r) - (e_p\star  e_q)\star  e_r=  e_q\star (e_p\star  e_r) - ( e_q\star  e_p)\star  e_r, \;\;p, q, r\in \mathbb Z,
\eea
and re-interpreted, in the language of $2-$cocycle, the property of a bilinear form to provide a central extension of a quasiassociative algebra. His study led to a complex on the space of cochains  and its generalization. Besides, Kuperschmidt discussed the homology and performed the differential-variational versions of the main results for the case when the centerless Virasoro algebra is replaced by the Lie algebra of vector fields on the circle.

This paper addresses a  generalization of the  algebra (\ref{virkuper}), denoted by $({\mathcal A, [. , .]}),$
endowed with  the multiplication
\beq\label{gvir}[e_{x_i}, e_{x_j}]=  g({x_i}, {x_j}) e_{x_i +x_j},\eeq
coming from the commutator
\beq\label{com}[e_{x_i}, e_{x_j}]= a e_{x_i}\star e_{x_j}- b e_{x_j}\star e_{x_i},\eeq
where $ (a,b)\in \mathbb R\times \mathbb R^+, \; ({x_i}, {x_j})\in \mathbb Z^{2}.$
We give the necessary and sufficient condition for this algebra to be
 a quasiassociative algebra with the multiplication
\beq\label{qas} e_{x_i}\star e_{x_j}= f({x_i}, {x_j}) e_{x_i +x_j}.
\eeq

Let us immediately mention  that such a generalization  of the  algebra (\ref{virkuper}) can lead to various classes of nonassociative algebras \cite{schafer} such as  alternative algebras, Jordan algebras, and so on, as well as to their various extensions, depending on the  defining  functions $f$ and $g$, but also on the real constants $a$ and $b.$ 
However, without loss of generality, in the sequel, the functions $f$ and $g$ are assumed to be defined as follows:

\bea
f, \, g: \mathbb Z \times \mathbb Z& \longrightarrow& \mathbb Z, \;\;(x_i, x_j ) \mapsto f(x_i, x_j ), \,g(x_i, x_j ).
\eea
Moreover, in this work, we are only interested in the class of left symmetric algebras, also called quasi-associative algebras.

The main results obtained in this work can be  summarized in the four following theorems:
\begin{theorem}
%\begin{enumerate}
%\item[(i)]
 For the multiplication 
 $e_{x_i}\star e_{x_j}:= R_{ij}^k e_{x_k},$ 
 the  quasi-associativity condition
\bea
(e_{x_i}\star e_{x_j})\star e_{x_l} - e_{x_i}\star (e_{x_j}\star e_{x_l}) = (e_{x_j}\star e_{x_i})\star e_{x_l} - e_{x_j}\star (e_{x_i}\star e_{x_l}),
\eea
 is expressed by  the Nijenhuis-torsion free relation:
\bea\label{nijenhuis}
\Big(R_{ij}^k - R_{ji}^k\Big)R_{kl}^m - R_{jl}^kR_{ik}^m+R_{il}^kR_{jk}^m=0.
\eea
\end{theorem}

\begin{theorem}
%\item[(ii)]
The hereditary condition for a linear map $\Phi: \mathcal A\, \longrightarrow \;\mathcal A$ associated with  the generalized Virasoro algebra (\ref{gvir}) makes into:
\begin{itemize}
\item For $\Phi: e_{x_i} \mapsto  e_{x_i+x_0}$ for some fixed  $x_0 \in \mathbb Z,$ $\forall x_i \in \mathbb Z,$ 
\beq\label{hc}
g(x_i+x_0, x_j) +g(x_i, x_j+x_0)-g(x_i, x_j)= g(x_i+x_0, x_j+x_0),
\eeq
or, equivalently,

\beq
\Big(\mathcal T_{x0} + \mathcal L_{x0} -\mathcal T_{x0}\mathcal L_{x0} - 1\Big) g(x_i, x_j)= 0,
\eeq
where the right and  left  translation operators $\mathcal T_{x0}$ and $\mathcal L_{x0}$ are defined, respectively,  by

\beq
\mathcal T_{x0}g(x_i, x_j):=g(x_i, x_j+x_0), \;\;\mathcal L_{x0}g(x_i, x_j):=g(x_i+x_0, x_j);
\eeq

\item For $\Phi: e_{x_i} \mapsto R_i^k e_{x_k},$
\bea
\Big[g(x_m,x_j)R_i^mR_{m+j}^s +g(x_i,x_n)R_j^nR_{n+i}^s -g(x_i,x_j)R_{i+j}^tR_t^s\Big]e_{x_s}= g(x_m, x_n)R_i^mR_j^n e_{x_m+x_n};\nonumber \\
\eea
\item For $\Phi: e_{x_i} \mapsto e_{x_i+1}, \; \; [e_{x_i}, e_{x_j}]:=R_{ij}^k e_{x_k},$
\bea
R_{(i+1)j}^k +R_{(j+1)i}^k-R_{ij}^k -R_{(i+1)(j+1)}^k= 0.
\eea
\end{itemize}
\end{theorem}
%\item[(iii)]

\begin{theorem}
For a map  $\rho: \, \mathcal A\, \longrightarrow \;\mathcal A, \, e_{x_i}\longmapsto e_{x_i+x_0}$ with  some fixed $x_0 \in \mathbb Z,$ $\forall x_i \in \mathbb Z,$ the $\rho-$compatibility equation  is equivalent to the invariance of the operator $(1-\mathcal E)$ under the action of the right translation operator $\mathcal T_{x_0},$ i.e.,
\beq\label{case0}
\mathcal T_{x_0}(1-\mathcal E)=(1-\mathcal E),
\eeq
where  the right translation and exchange operators $\mathcal T_{x_0}$ and $\mathcal E$ are defined,
respectively, by
\beq
\mathcal T_{x_0}f(x_i,x_j)= f(x_i, x_j+x_0), \;  \mathcal E f(x_i, x_j)= f(x_j,x_i).
\eeq
\end{theorem}

\begin{theorem}
%\item[(iv)] 
The universal identity, known for  nonassociative algebras \cite{sokolov2}, turns out to be in the following form for the  generalized   algebra introduced  with the multiplication (\ref{qas}):

\bea
&&f(x_k, x_j)\Big(-\mathcal L_{x_i}(x_j, x_k+x_j)(1-\mathcal E)(f(x_i, x_j) +[[f(x_j, x_k+x_s), f(x_i, x_k+x_j)]]\Big)\cr
&-&\mathcal L_{x_i+x_j}f(x_k,x_s)\Big(-\mathcal L_{x_j}(x_i, x_k)(1-\mathcal E)(f(x_i, x_j) +[[f(x_j, x_k), f(x_i, x_k)]]\Big)\cr
&-&\mathcal T_{x_i+x_j}f(x_k,x_j)\Big(-\mathcal L_{x_j}(x_i, x_j)(1-\mathcal E)(f(x_i, x_j) +[[f(x_j, x_s), f(x_i, x_s)]]\Big)=0
\eea

where
\beq
[[f(x_j, x_l), f(x_i, x_l)]]=f(x_j,x_l)\mathcal T_{x_s}f(x_i,x_l)-f(x_i,x_l)\mathcal T_{x_i}f(x_j,x_l),
\eeq
and $\mathcal L_u, \; \mathcal T_v$ are the usual left and right  translation operators, respectively.
%\end{enumerate} 
\end{theorem}

}

In the sequel, we give a full characterization of this generalized Virasoro algebra for which  the quasi-associativity condition and   the criteria  making it  a Lie  algebra   are discussed.  We deduce the hereditary operator and its generalization to the corresponding $3-$ary bracket. Further, we deduce the so-called $\rho-$compatibility equation, and investigate a  phase-space extension. Finally, concrete relevant particular situations  are analyzed.

The paper is organized as follows. Section $2$ deals with some preliminaries on left symmetric algebras (also called quasi-associative algebras). In section $3,$ we  discuss the main   properties of the generalized Virasoro algebra.  The case of left-alternative algebra structure and its link to some classes of nonlinear systems of differential equations are also recalled. Coboundary operators and  a $3-$ary bracket are defined and discussed. We define the hereditary operator,  and generalize it to the case of $3-$ary bracket.   Then, we derive the 
associated $\rho-$ compatibility equation. Phase-space extension is also discussed.
 In section $4$, we investigate the full centrally extended Virasoro algebra in the framework of the considered formalism. In section $5, $ we analyze the case of { an  infinite} dimensional Lie algebra of polynomial vector fields on the real line $\mathbb R^1$  and deduce some remarkable identities. We end, in section $6,$ by some concluding remarks.

\section{Preliminaries on left symmetric algebras}

The ordinary  centerless Virasoro-Witt algebra belongs to the class of quasi-associative algebras, known, 
in the literature \cite{bai} (and references therein), under the name of left-symmetric algebras (LSAs), arising in many areas of mathematics and physics. The LSAs were initially introduced by Caley in 1896 in the context of 
rooted tree algebras and in recent years Vinberg and Koszul re-introduced 
them in the context of convex homogeneous cones. LSAs { have }also independently appeared in the works by Gerstenhaber. 
As a consequence, perhaps, LSAs are known under different names. They are also called Vinberg algebras, 
Koszul algebras, or quasi-associative algebras, Gerstenhaber algebras, or pre-Lie algebras.
See \cite{burde}, \cite{wass} and \cite{winterhalder} (and references therein).
\smallskip

Geometrically, they are also connected to the theory of affine manifolds and  affine structures on Lie groups. See  \cite{auslander}, \cite{milnor},  \cite{burde} and references therein. Recall a smooth manifold which admits a linear connection $\triangledown$ whose torsion and curvature tensor vanish, is called an affinely flat (or simply affine in short) manifold. By a well known theorem of differential geometry, such a manifold is locally equivalent to an open subset of Euclidean space with the standard connection, i. e., for each point of the manifold, there are a neighborhood and a coordinate map into the Euclidean space which is an affine equivalence. 
In fact, the torsion and curvature are exactly the obstructions to the existence of such a map. In general, a connection $\triangledown$ on a Lie group is completely determined by the action on the left invariant vector fields, i.e., by $\triangledown_XY$ for $X,Y\in \frak g$ using the Leibniz rule.
 $\triangledown$ is left-invariant if and only if $\triangledown_XY\in \frak g$ whenever $X, Y\in \frak g.$ To perceive the problem algebraically, denote $\triangledown_XY$ by $X.Y$ for a left-invariant connection
$\triangledown$ and vector fields $X, Y\in \frak g$. Then having a left-invariant connection on $G$ is the same as having an algebra structure on $\frak g$. In this way, the geometric problems involving left-invariant connection become algebraic ones. 

A left-invariant connection $\triangledown$ on $G$ is said to be bi-invariant if it is also right-invariant. As usual, this holds if and only if $\triangledown$ is adjoint invariant. We can characterize bi-invariant connections using the associated algebra structure $X.Y= \triangledown_XY$ as follows  \cite{kim}:
\begin{proposition}
The following statements are equivalent:
\begin{itemize}
\item[(i)] A left-invariant connection  $\triangledown$ on $G$ is bi-invariant.
\item[(ii)] $\mbox{Ad}_g$ is an algebra automorphism on $(\frak g, .)$ for all $g \in G.$
\item[(iii)] $\mbox{Ad}_X$ is an algebra derivation on $(\frak g, .)$ for all $X\in \frak g,$ i.e.,
\bea
\mbox{Ad}_X(Y.Z)&=& \mbox{Ad}_X(Y).Z + Y. \mbox{Ad}_X(Z)\cr
\mbox{or}\;\; [X, Y.Z]&=& [X, Y]. Z + Y. [X, Z], X, Y, Z\in \frak g.
\eea
\end{itemize}
\end{proposition}
\proof See  \cite{kim}.
\qed

If furthermore the connection  $\triangledown$ is torsion free, then the algebra automorphism becomes a Lie algebra automorphism since $[X, Y]= X.Y - Y. X,\;\;\; X, Y \in \frak g$. Suppose further that  $\triangledown$ is affinely flat so that it has vanishing torsion and curvature tensor. Then,  using small letters for elements of $\frak g$ and $xy$  for $X.Y= \triangledown_XY$,  the torsion-free condition and the flatness of $\triangledown$ become algebraically:

\bea
xy-yx= [x, y] \label{lie}\\
x(yz)- y(xz)-[x,y]z=0,
\eea
respectively, 
for all $x, y , z\in \frak g.$ This leads to the following definition.

\begin{definition}
Let $A$ be a vector space over a field ${\mathbb K}$ equipped with a bilinear product $(x,y) \mapsto xy$.
$A$ is called a {\it left symmetric} algebra, {or, equivalently,  a {\it quasi-associative} algebra}, if for all $x,y,z \in A$ the associator 
 $\Big(x,y,z\Big)= x(yz) - (xy)z$ is symmetric in $x,y$, i.e., 
\bea\label{elsa}
\Big(x,y,z\Big)=\Big(y,x,z\Big), \qquad \hbox{ or} \qquad (xy)z - x(yz) = (yx)z - y(xz).
\eea
\end{definition}
Hence finding a left invariant  affinely flat connection on $G$ is the same as finding a left-symmetric algebra structure on $\frak g$ which is compatible  with  Lie algebra structure of $\frak g$ in the sense of (\ref{lie}).

Left symmetric algebras are Lie-admissible algebras (cf. \cite{medina}). Let $A$ be a LSA, then for any $x \in A$, denote
$L_x$ the left multiplication operator, $L_x(y) = xy$ for all $y \in A$. By setting $[x,y] := xy - yx$ a Lie bracket
defines a Lie algebra ${\mathcal G}(A)$, known as the sub-adjacent Lie algebra of $A$. Thus $A$ is called a compatible
left symmetric algebra structure on the Lie algebra ${\mathcal G}(A)$. 

Let ${\mathcal G}(A) \to gl(A)$ with $x \mapsto L_x$. Then $(L,A)$ gives a representation of the Lie algebra
${\mathcal G}(A)$, i.e., $[L_x,L_y] = L_{[x,y]}$ for all $x,y \in A$. But this is neither sufficient nor necessary
condition to give compatible LSA on any Lie algebra. These are given as follows. Let ${\mathcal G}$ be a Lie algebra
with a representation $\rho: {\mathcal G} \to gl(V)$, then a one-cocycle $q : {\mathcal G} \to V$ be a linear map
associated to $(\rho,q)$ such that $q[x,y] = \rho(x)q(y) - \rho(y)q(x)$ for all $x,y \in {\mathcal G}$. It has been shown in \cite{bai} that there is a compatible LSA structure if and only if there exists a bijective one-cocycle of ${\mathcal G}$. 

If $(\rho,q)$  is a bijective one-cocycle of ${\mathcal G}$ then
$x \ast y = q^{-1}\rho(x)q(y)$  defines a LSA structure on ${\mathcal G}$, where as for a LSA $A$ the identity transformation
is a one-cocycle of ${\mathcal G}(A)$ associated to the regular representation $L$.

\bigskip

It is also worth noticing that left-symmetric structures appear of natural way in the theory of integrable systems of hydrodynamic type (see \cite{kuper2}, \cite{dubrovin}, \cite{tsarev} (and references therein)) and the generalized Burgers equation.  Indeed, it was proved by Sokolov {\it and co-workers} \cite{habib}, (see also \cite{sokolov2} and references therein), the following theorem giving the link between LSAs and multicomponent generalizations of Burgers equations.

\begin{theorem}
If $C_{jk}^i$ are structure constants of any LSA, then the system of Burgers equations,
\bea
u_t^i= u_{xx}^i + 2 C_{jk}^i u^ku_x^j + A_{jkm}^i u^ku^ju^m, \;\;\mbox{where}\;\; i, j, k =1,\ldots, N
\eea
is integrable iff the following relations hold:

\bea
A_{jkm}^i&=& {1\over 3}\Big( C_{jr}^iC_{km}^r+C_{kr}^iC_{mj}^r+C_{mr}^iC_{jk}^r-C_{rj}^iC_{km}^r-C_{rk}^iC_{mj}^r-C_{rm}^iC_{jk}^r\Big)\cr
&&C_{jr}^iC_{km}^r-C_{kr}^iC_{jm}^r= C_{jk}^rC_{rm}^i-C_{kj}^rC_{rm}^i.
\eea
\end{theorem}

In this case, let $e_1, \ldots, e_N$ be a basis of a LSA $\mathcal A$, and $u=u^ie_i$. Then, the integrable system can be written as 

\bea
u_t=u_{xx} + 2 u\circ u_x + u\circ (u\circ u)-(u\circ u)\circ u,
\eea
where $\circ$ denotes the multiplication in $\mathcal A$.

\section{Generalized Virasoro algebra: quasi-associativity, hereditary operator and $\rho-$compatibility equation}

{ In this section, we  discuss the main   properties of the generalized Virasoro algebra.  The case of left-alternative algebra structure and its link to some classes of nonlinear systems of differential equations are also recalled.  A $3-$ary bracket is defined and the relation between the functions $f$ and $g$ is given. We define the hereditary operator,  and generalize it to the case of $3-$ary bracket.   Then, we derive the 
associated $\rho-$ compatibility equation}.

\subsection{Skew-symmetry, Jacobi identity, coboundary operators and derivation property}
\begin{itemize}
\item Skew-symmetry

The skew-symmetry property
\bea
 [e_{x_i}, e_{x_j}]&= &a e_{x_i}\star e_{x_j}- b e_{x_j}\star e_{x_i}=g({x_i}, {x_j}) e_{x_i +x_j}\cr
&=& -[e_{x_j}, e_{x_i}]= - \Big(a e_{x_j}\star e_{x_i}- b e_{x_i}\star e_{x_j}\Big)= - g({x_j}, {x_i}) e_{x_i +x_j}
\eea
induces the conditions{
\beq\label{sk}
a= b\; \mbox{or}\; g(x_i, x_j)= - g(x_j, x_i).
\eeq
}
\item Jacobi identity criterion

The Jacobi identity
\beq
\Big[ [e_{x_i}, e_{x_j}],e_{x_k}\Big] + \Big[ [e_{x_j}, e_{x_k}],e_{x_i}\Big] + \Big[ [e_{x_k}, e_{x_i}],e_{x_j}\Big]= 0
\eeq
reduces to a condition similar to the {\it Bianchi's  identity }
\beq\label{J11}
{\mathbb J}_{ij}^k + {\mathbb J}_{jk}^i + {\mathbb J}_{ki}^j=0
\eeq
where
\beq
{\mathbb J}_{ij}^k:= g({x_i}, {x_j}) g({x_i + x_j}, {x_k}). 
\eeq
Exploiting the relation 
%(\ref{fg})
 between the functions $f$ and $g,$ i.e.,

\beq\label{fg}
g(x_i, x_j)= af(x_i, x_j) - bf({x_j, x_i}).
\eeq

we can re-express the criterion for the Jacobi identity by the following result:
\bea\label{J2}
&& a^2\Big({\mathbb T}_{ij}^k + {\mathbb T}_{jk}^i + {\mathbb T}_{ki}^j\Big) + b^2 \Big({\mathbb G}_{ji}^k + {\mathbb G}_{ik}^j + {\mathbb G}_{kj}^i\Big)\cr
&-& ab\Big({\mathbb G}_{ij}^k + {\mathbb G}_{jk}^i + {\mathbb G}_{ki}^j 
+{\mathbb T}_{kj}^i + {\mathbb T}_{ji}^k + {\mathbb T}_{ik}^j\Big)=0
\eea
where
\beq
{\mathbb T}_{ij}^k := f({x_i}, {x_j}) f({x_i + x_j}, {x_k}),\;\;\;\;{\mathbb G}_{ij}^k =:f({x_i}, {x_j}) f({x_k,  x_i+x_j}).
\eeq

The criteria (\ref{sk}) and (\ref{J11}) or (\ref{J2}) confer the Lie algebra structure to the algebra $({\mathcal A}, [.,.])$. 

{ For $a=b,$ skew-symmetric functions  $f,$ (i. e. $f(x_i, x_j)= - f(x_j, x_i)$ for all $x_i, x_j \in \mathbb Z$), are  solutions of (\ref{J2}), which is consistent with the above skew-symmetry conditions (\ref{sk}). Note, however,  that the search for a general solution to this functional equation remains an open issue. }

\item Coboundary operators, $2-$cocycle and second cohomology

In addition to the above criteria, we consider an associated  Lie module ${\mathcal M}$ over ${\mathcal A},$ and a $k-$cochain, i.e. an alternating ${\mathbb K}-$multilinear map 
$$\psi:\,{\mathcal A}\times {\mathcal A}\times \ldots\times {\mathcal A} \,(k\,\mbox{copies of}\,{\mathcal A})  \longrightarrow {\mathcal M}.$$

The most important ${\mathcal A}$-modules are the trivial module ${\mathbb K}$, i.e., the action reads
$e_{x_n} \cdot {\lambda} = 0$ for all $\lambda \in {\mathbb K}$ and all $e_{x_n} \in {\mathcal A},$ and adjoint module
${\mathcal A}$, i.e., ${\mathcal A}$ acts on ${\mathcal A}$ by the adjoint action.

\smallskip

\noindent
Denote the vector space of $k-$cochains by ${\mathcal C}^k({\mathcal A}, {\mathcal A})$ and define the coboundary operators
\begin{eqnarray}
\delta_k: {\mathcal C}^k({\mathcal A}, {\mathcal A})\longrightarrow {\mathcal C}^{k+1}({\mathcal A}, {\mathcal A}),\, k\in \mathbb N, \, \mbox{with}\, \delta_{k+1}\circ \delta_k=0.
\end{eqnarray}
\begin{definition}
A $k-$cochain $\psi$ is called a $k-$cocycle if it lies in the kernel of the coboundary operator $\delta_k.$ It is called a $k-$coboundary if it lies in the image of the (k-1) coboundary operator. 
\end{definition}

A skew-symmetric map $\psi : {\mathcal A} \times {\mathcal A} \to {\mathcal A}$ is a Lie algebra $2-$cocycle
with values in the adjoint module if
\begin{eqnarray}\label{cohomology}
\delta_2\psi(e_{x_i}, e_{x_j}, e_{x_k}):&=& \psi([e_{x_i}, e_{x_j}], e_{x_k}) + \psi([e_{x_j}, e_{x_k}], e_{x_i}) + \psi([e_{x_k}, e_{x_i}], e_{x_j})\cr &-& [e_{x_i} , \psi(e_{x_j}, e_{x_k})]
+ [e_{x_j} , \psi(e_{x_i}, e_{x_k})] - [e_{x_k} , \psi(e_{x_i}, e_{x_j})] = 0
\end{eqnarray}

and a coboundary if there exits a linear map $\phi :{\mathcal A} \to {\mathcal A}$ with  
\begin{eqnarray}\label{coboundary}
\psi(e_{x_i}, e_{x_j}) = (\delta_1\phi)(e_{x_i}, e_{x_j}):=\phi([e_{x_i}, e_{x_j}]) - [e_{x_i},\phi( e_{x_j})] + [e_{x_j},\phi( e_{x_i})]. 
\end{eqnarray}
%A $k-$cochain $\psi$ is called a $k-$cocycle if it lies in the kernel of the coboundary operator $\delta_k.$ It is called a $k-$coboundary if it lies in the image of the (k-1) coboundary operator. 
The second cohomology of ${\mathcal A}$ with values in the adjoint representation is 
$$ H^2({\mathcal A}, {\mathcal A}) = Ker\, \delta_2 / Im\, \delta_1, $$ whereas $H^2({\mathcal A},{\mathbb K})$ with values in the trivial module
is related to the central extension of ${\mathcal A}$. It is worth to say that deformations of the Lie algebra ${\mathcal A}$ are related to the
Lie algebra cohomology and $H^2({\mathcal A}, {\mathcal A})$ classifies infinitesimal deformations \cite{Gerst}.
If $H^2({\mathcal A}, {\mathcal A}) = 0$, then ${\mathcal A}$ is 
infinitesimally formally rigid.

An elementary and direct calculation of the  vanishing second Lie algebra cohomology 
of the Witt and  Virasoro algebras with values in the adjoint module is given by  Schlichenmaier \cite{Schli}.
In 1989, A. Fialowski showed by explicit calculations the vanishing of the second Lie algebra cohomology of the Witt algebra 
(in an unpublished manuscript). She also gave statements of the rigidity of the Witt and Virasoro algebras \cite{Fial} without proof.

\smallskip

Write  the $0-$cocycle as 
\begin{eqnarray}
\psi(e_{x_i}, e_{x_j})= \psi_{x_i,x_j}e_{x_i+x_j}.
\end{eqnarray}
If it is a coboundary, then it can be given as a coboundary of a linear form of degree $0$, i.e., 
\begin{eqnarray}
\phi(e_{x_i})= \phi_{x_i}e_{x_i}.
\end{eqnarray}
Then the following result holds.
\begin{proposition}
The $2-$cocycle $\psi$ defined by
\begin{eqnarray}
\delta_2\psi(e_{x_i}, e_{x_j}, e_{x_k}):=0
\end{eqnarray}
leads to the functional relation:
\begin{eqnarray}
&& g(x_i, x_j)\psi_{x_i+x_j, x_k} + g(x_j, x_k)\psi_{x_j+x_k, x_i} + g(x_k, x_i)\psi_{x_i+x_k, x_j}\cr
& -& g(x_i, x_j+x_k)\psi_{x_j, x_k} +  g(x_j, x_i+x_k)\psi_{x_i, x_k} -  g(x_k, x_i+x_j)\psi_{x_i, x_j}= 0.
\end{eqnarray}
 Besides, there results from the  expression (\ref{coboundary}):
\begin{eqnarray}
(\delta_1 \phi)({x_i, x_j})= g(x_i, x_j) \left(\phi_{x_i+x_j} - \phi_{x_i} - \phi_{ x_j}\right)e_{x_i + x_j}.
\end{eqnarray}
Hence, $\psi$ is a coboundary if and only if there exists a system of $\phi_{x_k}\in \mathbb C$, $x_k\in \mathbb Z,$ such that
\begin{eqnarray}
\psi_{x_i, x_j}= g(x_i, x_j) \left(\phi_{x_i+ x_j} - \phi_{x_i} - \phi_{ x_j}\right).
\end{eqnarray}
\end{proposition}
\end{itemize}

\begin{itemize}
\item Derivation property

This property expressed as
\beq
[e_{x_i}, e_{x_j}\star e_{x_k}]: = e_{x_j}\star [e_{x_i},  e_{x_k}] +[e_{x_i}, e_{x_j}]\star e_{x_k}
\eeq
leads to
\beq
f({x_j}, {x_k}) g({x_i,  x_j+x_k})=g(x_i, x_k)f({x_j,  x_i+x_k}) + g(x_i, x_j)f({x_i + x_j,  x_k}) 
\eeq
or, equivalently, 
using the relation (\ref{fg}) between $f$ and $g,$ to
\bea
&&f({x_j}, {x_k})\Big[af(x_i, x_j+x_k) - bf({x_j+x_k, x_i})\Big] \cr
&-&f({x_j}, {x_i+x_k}) 
\Big[af(x_i, x_k) - bf({x_k, x_i})\Big] 
-f({x_i+x_j}, {x_k})\Big[af(x_i, x_j) - bf({x_j, x_i})\Big]=0 \nonumber \\
\eea
which can be simply rewritten as
\beq\label{der}
a\Big({\mathbb G}_{jk}^i  - {\mathbb G}_{ik}^j  -{\mathbb T}_{ij}^k \Big) + b
\Big({\mathbb G}_{ki}^j  + {\mathbb T}_{ji}^k  -{\mathbb T}_{jk}^i\Big)= 0.
\eeq
\end{itemize}

Let us extend now the notion of skew-symmetry and Jacobi identity to   the vector space ${\mathcal B}= {\mathcal A}\oplus {\mathcal A}$ with the multiplication
\beq
\Big[(e_{x_p},e_{x_r}), (e_{x_q},e_{x_s})\Big]:=\Big([e_{x_p},e_{x_q}], e_{x_p}\star e_{x_s} - e_{x_q}\star e_{x_r}\Big)
\eeq

where $(.,.)$ is the ordinary dot product. Explicitly, this new bracket gives

\beq
\Big[(e_{x_p},e_{x_r}), (e_{x_q},e_{x_s})\Big]= g(x_p,x_q)\Big(e_{x_p+x_q}, f(x_p,x_s)e_{x_p+x_s} -
f(x_q,x_r)e_{x_q+x_r}\Big),
\eeq

or, equivalently,

\beq
\Big[(e_{x_p},e_{x_r}), (e_{x_q},e_{x_s})\Big]= g(x_p,x_q)\Big[f(x_p,x_s)\Big(e_{x_p+x_q}, e_{x_p+x_s}\Big) -
f(x_q,x_r)\Big(e_{x_p+x_q},e_{x_q+x_r}\Big)\Big].
\eeq
Therefore, the  skew-symmetry criterion reads
\beq
\Big[g(x_p,x_q)-g(x_q,x_p)\Big]\Big[f(x_p,x_s)\Big(e_{x_p+x_q},e_{x_p+x_s}\Big)-
f(x_q,x_r)\Big(e_{x_p+x_q},e_{x_q+x_r}\Big)\Big]=0.
\eeq
The Jacobi identity property
\bea
\Big[[(e_{x_p},e_{x_r}), (e_{x_q},e_{x_s})], (e_{x_t},e_{x_u})\Big] &+& \Big[[(e_{x_q},e_{x_s}), (e_{x_t},e_{x_u})], (e_{x_p},e_{x_r})\Big]\cr &+&\Big[[(e_{x_t},e_{x_u}), (e_{x_p},e_{x_r})], (e_{x_q},e_{x_s})\Big]=0
\eea
gives here the following functional relation:
\bea
\Big[ g(x_p,x_q)g(x_p + x_q, x_t) f(x_p+x_q, x_u)
- g(x_q,x_t)g(x_q + x_t, x_p) f(x_q,x_u)f( x_p, x_q+x_u)\nonumber\\
+ g(x_t,x_p)g(x_t + x_p, x_q) f(x_p,x_u) f(x_q,x_p+x_u)\Big]\Big(e_{x_p+x_q+x_t},e_{x_p+x_q+x_u} \Big)
\nonumber\\
-\Big[ g(x_p,x_q)g(x_p + x_q, x_t) f(x_p,x_s) f(x_t, x_p+x_s)
%\right.\nonumber\\
- 
%\left. 
g(x_q,x_t)g(x_q + x_t, x_p) f(x_t,x_s)f( x_p, x_t+x_s)\nonumber\\
- g(x_t,x_p)g(x_t + x_p, x_q) f(x_p+x_t,x_s)\Big]\Big(e_{x_p+x_q+x_t},e_{x_p+x_t+x_s} \Big)\nonumber\\
+\Big[ g(x_p,x_q)g(x_p + x_q, x_t) f(x_q,x_r) f(x_t, x_q+x_r)
%\right.\nonumber\\
+
%\left. 
g(x_q,x_t)g(x_q + x_t, x_p) f(x_q+x_t,x_r)\nonumber\\
- g(x_t,x_p)g(x_t + x_p, x_q) f(x_t,x_r)f(x_q,x_r+x_t)\Big]\Big(e_{x_p+x_q+x_t},e_{x_q+x_r+x_t} \Big)=0,\nonumber
 \eea
 or, equivalently,

\bea
&&\Big[\mathbb J_{qp}^t\mathcal L_{x_q} f(x_p, x_u) - \mathbb J_{qt}^p\mathbb G_{qu}^p +\mathbb J_{tp}^q\mathbb G_{pu}^q\Big]\Big(e_{x_p+x_q+x_t},e_{x_p+x_q+x_u} \Big)\cr
&+&\Big[ \mathbb J_{tp}^q\mathcal L_{x_t} f(x_p, x_s) + \mathbb J_{qt}^p\mathbb G_{ts}^p- \mathbb J_{pq}^t\mathbb G_{ps}^t \Big]
\Big(e_{x_p+x_q+x_t},e_{x_p+x_t+x_s} \Big)\cr
&+&\Big[ \mathbb J_{pq}^t\mathbb G_{qr}^t + \mathbb J_{qt}^p\mathcal L_{x_q} f(x_t, x_r) - \mathbb J_{tp}^q\mathbb G_{tr}^q \Big]\Big(e_{x_p+x_q+x_t},e_{x_q+x_r+x_t} \Big)= 0,
\eea

where $\mathcal L_{x_t} $ stands for the left-translation operator acting as $\mathcal L_{x_t} f(x_q, x_r)=f(x_q+x_t, x_r).$

%The properties  (\ref{sk}) and (\ref{J1}) or (\ref{J2}) together with  (\ref{der}) make $({\mathcal A}, [.,.])$ into a Poisson algebra.
%\end{itemize}

\subsection{Quasi-associativity condition}
From the definition of the quasiassociativity w.r.t. the multiplication rule (\ref{qas}), we infer the following relation 
\beq\label{qasc}
\Big[f({x_i}, {x_j}) -f({x_j}, {x_i})\Big]f({x_i+x_j, x_k})= f({x_j}, {x_k})f({x_i}, {x_j+x_k}) -   f({x_i}, {x_k})f({x_j}, {x_i+x_k})
\eeq
expressing the necessary and sufficient property which confers a space phase structure to the sub-adjacent Lie algebra\cite{kuper1}.

Setting { the ansatz}
\beq
f({x_i}, {x_j}):= \phi_{x_i}(x_j)
\eeq
for some linear map { $\phi: {\mathbb Z}\rightarrow \mbox{End}(\mathbb Z),\, \phi_x= \phi(x),$} then the quasiassociativity condition (\ref{qasc}) reads
\beq
\Big[\phi_{x_i}(x_j) - \phi_{x_j}(x_i)\Big]\phi_{x_i+x_j}(x_k)=\phi_{x_j}(x_k)\phi_{x_i}(x_j+x_k) - \phi_{x_i}(x_k)\phi_{x_j}(x_i+x_k),
\eeq
which can be further simplified into the expression 
%Simplifying further by defining $\phi$ such that   leads 
\beq\label{requasi}
\Big[ {\mathcal R}_i^j -  {\mathcal R}_j^i\Big] {\mathcal R}_{i+j}^k -  {\mathcal R}_j^k
 {\mathcal R}_i^{j+k} + {\mathcal R}_i^k {\mathcal R}_j^{i+k} = 0
\eeq
by defining $\phi_{x_i}(x_j):= {\mathcal R}_i^j.$
%This relation is  trivially satisfied, $0=0,$ for $i= j$,  i.e., in the case of left-alternative algebra in the sense of Kazakov \cite{kazakov}, $(aa)b=a(ab).$ 
The relation (\ref{requasi}) can be given an interpretation in terms of the null value of the 
Nijenhuis-torsion. Indeed, we have the following result.

\begin{proposition}
Let  $e_{x_i}\star e_{x_j}:= R_{ij}^k e_{x_k}.$ 
Then the  quasi-associativity condition, i.e.,
\bea
(e_{x_i}\star e_{x_j})\star e_{x_l} - e_{x_i}\star (e_{x_j}\star e_{x_l}) = (e_{x_j}\star e_{x_i})\star e_{x_l} - e_{x_j}\star (e_{x_i}\star e_{x_l}),
\eea
 is expressed by  the Nijenhuis-torsion free relation:
\bea\label{nijenhuis1}
\Big(R_{ij}^k - R_{ji}^k\Big)R_{kl}^m - R_{jl}^kR_{ik}^m+R_{il}^kR_{jk}^m=0.
\eea
\end{proposition}
This well agrees with the result exposed in \cite{winterhalder}. In this case, as well pointed out by Kuperschmidt \cite{kuper2} (see also references therein), all associated hydrodynamic systems are diagonalizable in Riemann invariants whenever they are hyperbolic. Moreover, a diagonal $N-$component hyperbolic hydrodynamic system whose Nijenhuis torsion is zero, is isomorphic to $N$ noninteracting scalar equations. 

{
\begin{definition}[Alternative algebra]\label{alternative}
An alternative algebra $\mathcal U$ over a field $\mathbb F$ is an algebra  defined with the two identities \cite{schafer}
\bea\label{lalt10}
x^2y=x(xy)\,\,\mbox{for all}\,\,x, y \in \mathcal U
\eea
and
\bea\label{lalt20}
yx^2=(yx)x\,\,\mbox{for all}\,\,x, y \in \mathcal U,
\eea
known, respectively, as left and right alternative laws.
\end{definition}
In terms of associators, (\ref{lalt10}) and (\ref{lalt20}) are equivalent to 
\bea\label{lalt1}
(x,x,y)= 0\,\,\mbox{for all}\,\,x, y \in \mathcal U
\eea
and
\bea\label{lalt2}
(y,x,x)= 0\,\,\mbox{for all}\,\,x, y \in \mathcal U,
\eea
respectively.}
Therefore, the relations (\ref{requasi}) and (\ref{nijenhuis1})   are  trivially satisfied, $0=0,$ for $i= j$,  i.e., in the case of left-alternative algebra.  
 %$(aa)b=a(ab).$ 
In accordance with the modified Riccati scheme introduced by Kazakov \cite{kazakov}, such a left-alternative algebra is associated with the following vector system of equations:
\beq\label{veq}
{\bf Y}_t= -{\bf Y}\star{\bf Y} + \epsilon,
\eeq 
where the vector function $\epsilon(t)$ takes value in the considered left alternative algebra (LAA). The nonlinear system (\ref{veq}) can be reduced to linear problems by means of this LAA. Notice that, in contrast to systems of hydrodynamic type, the nonlinear systems of equations generated by left-alternative algebras do not, in general, have integrals of the motion. For more details, see \cite{kazakov}. { It is also worth mentioning the well-known  Burgers vectorial equation \cite{sokolov2}:
\bea
{\bf u}_t= {\bf u}_{xx} + 2 {\bf u\star u}_x + {\bf u}\star({\bf u}\star{\bf u}) - ({\bf u}\star{\bf u})\star{\bf u},
\eea
which is one of the important examples of equations associated with LSAs. In the case of  the generalized Virasoro algebra examined in this work, the $\star-$ product is defined by the multiplication law (\ref{qas}). New examples of nonlinear systems may exist, but their full investigations remain totally open and may be  the core of our  forthcoming works.}

\subsection{$3-$ary bracket}

{ Ternary algebra  plays an important role in the construction of the world volume theories of
multiple M2 branes  \cite{bagger}. The ternary bracket was introduced by Nambu \cite{nambu} and developed by Filippov  \cite{filippov}. Several authors  \cite{curtright1}   studied Kac-Moody and 
centerless Virasoro (or Virasoro-Witt) $3-$algebras
and demonstrated some of their applications to the Bagger-Lambert-Gustavsson theory.  The su(1,1) enveloping algebra was used by Curtright {\it et al}  \cite{curtright} to construct 
ternary Virasoro-Witt algebra. Motivated by this work, we study the ternary algebra
of the generalized Virasoro algebra defined by the following  $3-$ary bracket:}

\beq\label{3brac}
[e_{x_i}, e_{x_j},e_{x_k}]:= e_{x_i}\star [e_{x_j},e_{x_k}] + e_{x_j}\star [e_{x_k},e_{x_i}] + e_{x_k}\star [e_{x_i},e_{x_j}].
\eeq
Putting
\beq
{\mathbb E}_{ij}^k:=g(x_i, x_j)f({x_k}, {x_i+x_j})e_{x_i+x_j+x_k}
\eeq
yields the following expression for the $3-$ary bracket:
\beq\label{3bracket0}
[e_{x_i}, e_{x_j},e_{x_k}]= {\mathbb E}_{ij}^k + {\mathbb E}_{jk}^i + {\mathbb E}_{ki}^j.
\eeq
In the other hand, using the relations (\ref{com}) and (\ref{qas}) the same  bracket can also be evaluated in terms 
of the real $a$ and $b.$  In this case, denoting by
\beq
{\mathbb F}_{ij}^k:=\Big [af(x_i, x_j) - bf({x_j, x_i})\Big]f({x_k, x_i+x_j})e_{x_i+x_j+x_k},
\eeq
we obtain
\beq\label{3bracket00}
[e_{x_i}, e_{x_j},e_{x_k}]= {\mathbb F}_{ij}^k + {\mathbb F}_{jk}^i + {\mathbb F}_{ki}^j.
\eeq
{ 
The relations (\ref{3bracket0}) and (\ref{3bracket00}) can therefore be  used to define a $3-$algebra generalization of the  generalized Virasoro algebra proposed by
Kupershmidt \cite{kuper1}. Indeed, such a formulation of the $3-$ary bracket, cyclic in the indices $i, j, k,$ reminiscent of the {\it Bianchi's identity} for curvature tensor in differential geometry, is quite relevant for detailed analysis of $3-$algebras as  will be developed in a forthcoming paper}.

\subsection{Hereditary operator}

The hereditary operator is defined as follows \cite{fuchssteiner2}:
\begin{definition} [Hereditary operator]

A linear map $\Phi
: ({\mathcal A}, \circ) \rightarrow  ({\mathcal A}, \circ)$  defined as

\beq
[a, b]_\Phi:=  (\Phi a)\circ b + a \circ(\Phi b)  - \Phi (a\circ b),
\eeq
where $\circ$ is some binary bilinear operator on ${\mathcal A},$ is called hereditary if 
\beq\label{her1}
\Phi[a, b]_\Phi= (\Phi a)\circ (\Phi b),
\eeq
or, equivalently,

\beq\label{her2}
\Phi^2(a\circ b) + (\Phi a)\circ (\Phi b)= \Phi\Big[(\Phi a)\circ b +a\circ  (\Phi b)\Big].
\eeq
\end{definition}

Next, let us introduce the map ${\mathcal L}_k$ such that

\beq
{\mathcal L}_k(\Phi)(b):= k\circ \Phi(b) -  \Phi(b) \circ k - \Phi(k\circ b)+\Phi(b\circ k)
\eeq
for all $b\in{\mathcal A}$. Then $\Phi$ is $k-$invariant iff ${\mathcal L}_k \Phi= 0.$
%\end{definition}, 

Formally, let us write
\beq
{\mathcal L}_a b = a\circ b -b\circ a.
\eeq
Then, the Leibniz rule reads
\beq
{\mathcal L}_a(\Phi\circ b) = {\mathcal L}_a(\Phi)\circ b +\Phi({\mathcal L}_ab).
\eeq

Therefore, the map $\Phi$ is hereditary iff
\beq
\Phi({\mathcal L}_a\Phi)= {\mathcal L}_{\Phi(a)}\Phi.
\eeq
%\begin{remark}
% A  hereditary $\Phi$  is equivalent to the Nijenhuis operator ${\mathcal N}.$
%\end{remark}

Three particular cases give rise to interesting simpler conditions as follows.

\begin{proposition}
The hereditary condition for  the generalized Virasoro algebra (\ref{gvir}) makes into:
\begin{itemize}
\item[(i)] For $\Phi: e_{x_i} \mapsto  e_{x_i+x_0}$ for some fixed $x_0,$
\beq\label{hc}
g(x_i+x_0, x_j) +g(x_i, x_j+x_0)-g(x_i, x_j)= g(x_i+x_0, x_j+x_0),
\eeq
or, equivalently,

\beq
\Big(\mathcal T_{x0} + \mathcal L_{x0} -\mathcal T_{x0}\mathcal L_{x0} - 1\Big) g(x_i, x_j)= 0,
\eeq
where the left and right translation operators $\mathcal T_{x0}$ and $\mathcal L_{x0}$ are defined by

\beq
\mathcal T_{x0}g(x_i, x_j):=g(x_i, x_j+x_0), \;\;\mathcal L_{x0}g(x_i, x_j):=g(x_i+x_0, x_j);
\eeq

\item[(ii)]  For $\Phi: e_{x_i} \mapsto R_i^k e_{x_k},$
\bea
\Big[g(x_m,x_j)R_i^mR_{m+j}^s +g(x_i,x_n)R_j^nR_{n+i}^s -g(x_i,x_j)R_{i+j}^tR_t^s\Big]e_{x_s}= g(x_m, x_n)R_i^mR_j^n e_{x_m+x_n};\nonumber \\
\eea
\item[(iii)]  For $\Phi: e_{x_i} \mapsto e_{x_i+1}, \; \; [e_{x_i}, e_{x_j}]:=R_{ij}^k e_{x_k},$
\bea
R_{(i+1)j}^k +R_{(j+1)i}^k-R_{ij}^k -R_{(i+1)(j+1)}^k= 0.
\eea
\end{itemize}
\end{proposition}

%As matter of simple illustration, let us assume $\Phi(e_{x_i})= e_{x_i+x_0},$ for some $x_0 \in \mathbb Z.$ Then the hereditarity condition for the generalized Virasoro algebra considered in this work can be expressed as
%\beq\label{hc}
%g(x_i+x_0, x_j) +g(x_i, x_j+x_0)-g(x_i, x_j)= g(x_i+x_0, x_j+x_0).
%\eeq
%introducing the left and right translation operators $\mathcal T_{x0}$ and $\mathcal L_{x0}$ such that
%\beq
%\mathcal T_{x0}g(x_i, x_j)=g(x_i, x_j+x_0), \;\;\mathcal L_{x0}g(x_i, x_j)=g(x_i+x_0, x_j),
%\eeq
%then, the condition (\ref{hc}) can be re-expressed in the following operator form:
%\beq
%\Big(\mathcal T_{x0} + \mathcal L_{x0} -\mathcal T_{x0}\mathcal L_{x0} - 1\Big) g(x_i, x_j)= 0.
%\eeq
%where $\mathbb 1$ and $\mathbb 0$ are the identity and null operators, respectively.
Besides, defining the action of $\Phi$ on a  $3-$ary bracket in the simple case of $\Phi: e_{x_i} \mapsto e_{x_i+x_0}$ as:
\beq
\Phi\Big[e_{x_i}, e_{x_j},e_{x_k}\Big]_\Phi:=\Big[\Phi (e_{x_i}), \Phi(e_{x_j}),\Phi(e_{x_k})\Big]
\eeq
yields a Bianchi like identity
\beq
 {\mathbb P}_{ij}^k + {\mathbb P}_{jk}^i + {\mathbb P}_{ki}^j= 0
\eeq
where 
\beq
 {\mathbb P}_{ij}^k :=\Big[g(x_i+x_0, x_j) +g(x_i, x_j+x_0)-g(x_i, x_j)- g(x_i+x_0,x_j+x_0)\Big]f(x_k+x_0,x_i+x_j+2x_0)
\eeq
or, equivalently, in operator form,
\bea
 {\mathbb P}_{ij}^k :=\Big(\mathcal T_{x0} + \mathcal L_{x0} -\mathcal T_{x0}\mathcal L_{x0} - 1\Big) g(x_i, x_j) \mathcal T_{2x0+x_j}\mathcal L_{x0}f(x_k,x_i).
\eea
Hereditary operators play an important role in the field of nonlinear evolution equations. Indeed, as showed in \cite{fuchssteiner},  they  generate on a systematic level many new classes of  nonlinear dynamical systems which possess infinite dimensional abelian groups of symmetry transformations. Their so-called permanence properties, given  by Fuchssteiner \cite{fuchssteiner}, allow to construct new hereditary operators out of given ones.

\subsection{$\rho-$compatibility equation}
Following \cite{kuper1}, define a new multiplication 
%by a perturbation of the previous left-symmetric one as follows:
\beq\label{starnew1}
e_{x_i}\star\rq{}e_{x_j}=e_{x_i}\star e_{x_j} +\epsilon h_{e_{x_i}, e_{x_j}}, \; \;\epsilon^2=0
\eeq
with some (bi-)linear map $h: {\mathcal A}\otimes{\mathcal A}\rightarrow {\mathcal A},$ such that it makes ${\mathcal A}$ into a left-symmetric algebra as well.
The new associator $(e_{x_i},e_{x_j},e_{x_k})_{\star\rq{}}$ can be expressed in terms of old one as follows:
\beq\label{qas2}
(e_{x_i},e_{x_j},e_{x_k})_{\star\rq{}}=(e_{x_i},e_{x_j},e_{x_k})_{\star}+\epsilon T\rq{}_{e_{x_i},e_{x_j},e_{x_k}}
\eeq
where
\beq
T\rq{}_{e_{x_i},e_{x_j},e_{x_k}}= h_{e_{x_i},e_{x_j}\star e_{x_k}}- h_{e_{x_i}\star e_{x_j,} e_{x_k}}+e_{x_i}\star h_{e_{x_j},e_{x_k}} - h_{e_{x_i},e_{x_j}}\star e_{x_k}.
\eeq
Therefore,  the left symmetric condition for  $({\mathcal A}, \star\rq{})$ is provided by the relation
\bea
T_{e_{x_i},e_{x_j},e_{x_k}}&= &T\rq{}_{e_{x_i},e_{x_j},e_{x_k}} - T\rq{}_{e_{x_j},e_{x_i},e_{x_k}}\cr
&=&h_{e_{x_i},e_{x_j}\star e_{x_k}} - h_{e_{x_j},e_{x_i}\star e_{x_k}}- h_{[e_{x_i},e_{x_j}], e_{x_k}}\cr
&+&e_{x_i}\star h_{e_{x_j},e_{x_k}} - e_{x_j}\star h_{e_{x_i},e_{x_k}}\cr
& -&\Big (h_{e_{x_i},e_{x_j}}-h_{e_{x_j},e_{x_i}}\Big)\star e_{x_k}=0.
\eea
Write here also
\beq\label{starnew2}
h_{e_{x_i},e_{x_j}}:=\phi_{e_{x_i}}(e_{x_j})
\eeq
for some linear map $\phi:{\mathcal A}\rightarrow {\mathcal A},\;\; \phi_{e_{x_i}}=\phi({e_{x_i}}).$

Then the quasiassociativity condition (\ref{qas2}) takes the following form:
\bea\label{qas3}
&&\phi_{e_{x_i}}(e_{x_j}\star e_{x_k})-\phi_{e_{x_j}}(e_{x_i}\star e_{x_k})-\phi_{[e_{x_i},e_{x_j}]}(e_{x_k}) +  e_{x_i}\star\phi_{e_{x_j}}(e_{x_k})\cr
&-& e_{x_j}\star\phi_{e_{x_i}}(e_{x_k})-[\phi_{e_{x_i}}(e_{x_j})-\phi_{e_{x_j}}(e_{x_i})]\star e_{x_k}=0.
\eea
The relation (\ref{qas3}) can be rewritten in an operator form as:
\beq\label{qas3n}
\left[\phi_{e_{x_i}}, L_ {e_{x_j}}  \right] - \left[\phi_{e_{x_j}}, L_ {e_{x_i}}  \right]= L_{\phi_{e_{x_i}}(e_{x_j})-\phi_{e_{x_j}}(e_{x_i})+\phi_{[e_{x_i},e_{x_j}]}}.
\eeq
Suppose now that $\phi_{e_{x_i}}$ is of a special form:
\beq\label{starnew3}
\phi_{e_{x_i}}=L_{\rho(e_{x_i})}, \;\;\rho\in \mbox{End}({\mathcal A}),
\eeq
with some operator $\rho: {\mathcal A}\rightarrow {\mathcal A}.$ Since for LSAs,
\beq
[L_u, L_v]= L_{[u,v]},\;\;\forall u, v \in  {\mathcal A}
\eeq
the equation (\ref{qas3}) becomes
\beq\label{wce}
L_{E(e_{x_i},e_{x_j})}= 0
\eeq
where
\bea
E(e_{x_i},e_{x_j})&=&[\rho(e_{x_i}), e_{x_j}]-[\rho(e_{x_j}), e_{x_i}]-\Big(\rho(e_{x_i})\star e_{x_j}- \rho(e_{x_j})\star e_{x_i} \Big) - \rho\Big( [e_{x_i}, e_{x_j}]\Big)\cr
&=& -e_{x_j}\star \rho(e_{x_i}) + e_{x_i}\star \rho(e_{x_j}) - \rho\Big(e_{x_i}\star e_{x_j}-e_{x_j}\star e_{x_i}\Big)
\eea
The $\rho-$compatibility equation then reads
\bea\label{comp}
E(e_{x_i},e_{x_j})= 0 \Leftrightarrow e_{x_i}\star \rho(e_{x_j}) - e_{x_j}\star \rho(e_{x_i})&=&   \rho\Big(e_{x_i}\star e_{x_j}-e_{x_j}\star e_{x_i}\Big)\cr
&:=&\rho\Big[f(x_i,x_j)e_{x_i+x_j}-f(x_j,x_i)e_{x_i+x_j}\Big],
\eea
instead of the weaker deformation condition (\ref{wce}). The operator $\rho$ is called a strong deformation.

For the particular case, when $\rho: e_{x_i}\longmapsto e_{x_i+x_0}$ for fixed $x_0 \in \mathbb Z,$ the $\rho-$compatibility equation (\ref{comp}) turns out to be a simpler difference equation
\beq\label{comps}
\Big[f(x_i,x_j+x_0)- f(x_i,x_j)\Big] - \Big[f(x_j,x_i+x_0)- f(x_j,x_i)\Big]= 0.
\eeq
Define the right translation and exchange operators $\mathcal T_{x_0}$ and $\mathcal E,$
respectively, by
\beq
\mathcal T_{x_0}f(x_i,x_j)= f(x_i, x_j+x_0), \;  \mathcal E f(x_i, x_j)= f(x_j,x_i).
\eeq

Then, the relation (\ref{comps}) reads
\beq
\mathcal T_{x_0}(1-\mathcal E) f(x_i,x_j)=(1-\mathcal E) f(x_i,x_j).
\eeq
Therefore the following result holds.
\begin{proposition}
Let $\rho: e_{x_i}\longmapsto e_{x_i+x_0}$ for  some fixed $x_0 \in \mathbb Z.$ Then the $\rho-$compatibility equation (\ref{comp}) is equivalent to the invariance of the operator $(1-\mathcal E)$ under the action of the right translation operator $\mathcal T_{x_0},$ i.e.,
\beq\label{case0}
\mathcal T_{x_0}(1-\mathcal E)=(1-\mathcal E) .
\eeq
\end{proposition}

\subsection{Universal identity}
All nonassociative algebras naturally arising in connection with integrable systems satisfy a universal identity \cite{sokolov2}, i.e., 
\beq\label{unidentity}
[[e_{x_i}, e_{x_j}, e_{x_k}\star e_{x_s}]]-[[e_{x_i}, e_{x_j}, e_{x_k}]]\star e_{x_s} -e_{x_k}\star [[e_{x_i}, e_{x_j}, e_{x_s}]]=0,
\eeq
where
\bea
[[x,y,z]]= (x,y,z)-(y,x,z).
\eea
Here we get
\bea
[[e_{x_i}, e_{x_j}, e_{x_k}\star e_{x_s}]]&=&f(x_k,x_s)\Big(\Big [f(x_j,x_k +x_s)f(x_i,x_j+x_k+x_s) \cr
&- &f(x_i,x_k +x_s)
  f(x_j,x_i+x_k+x_s)\Big] \cr
%\right.
%&\cr
&-&
%\left.
 f(x_i+x_j, x_k+x_s)\Big[f(x_i,x_j) - f(x_j,x_i)\Big]\Big)
\eea

\bea
[[e_{x_i}, e_{x_j}, e_{x_k}]]\star e_{x_s}&=&f(x_i+x_j+x_k, x_s)\Big( f(x_i+x_j,x_k)\Big[f(x_j,x_i)
%\right.
-
%\left.
 f(x_i,x_j)\Big]\cr &+& f(x_j,x_k)f(x_i,x_j+x_k)-f(x_i,x_k)f(x_j,x_i+x_k)\Big)
%\cr
\eea

\bea
e_{x_k}\star [[e_{x_i}, e_{x_j}, e_{x_s}]]&=&f( x_k, x_i+x_j+x_s)
%&\times&
\Big( f(x_i+x_j,x_s)\Big[f(x_j,x_i)
- f(x_i,x_j)\Big]\cr &+& f(x_j,x_s)f(x_i,x_j+x_s)
-
%\left.
f(x_i,x_s)f(x_j,x_i+x_s)\Big)=0.
\eea 
%transforming (\ref{unidentity}) into  the following operator form:
Therefore,

\begin{proposition}
{The universal identity (\ref{unidentity}) turns out to be of the following form for the generalized  quasiassociative algebra defined  with the multiplication (\ref{qas})}: 

\bea\label{function relation}
&&f(x_k, x_j)\Big(-\mathcal L_{x_i}(x_j, x_k+x_j)(1-\mathcal E)(f(x_i, x_j) +[[f(x_j, x_k+x_s), f(x_i, x_k+x_j)]]\Big)\cr
&-&\mathcal L_{x_i+x_j}f(x_k,x_s)\Big(-\mathcal L_{x_j}(x_i, x_k)(1-\mathcal E)(f(x_i, x_j) +[[f(x_j, x_k), f(x_i, x_k)]]\Big)\cr
&-&\mathcal T_{x_i+x_j}f(x_k,x_j)\Big(-\mathcal L_{x_j}(x_i, x_j)(1-\mathcal E)(f(x_i, x_j) +[[f(x_j, x_s), f(x_i, x_s)]]\Big)=0
\eea

where
\beq
[[f(x_j, x_l), f(x_i, x_l)]]=f(x_j,x_l)\mathcal T_{x_s}f(x_i,x_l)-f(x_i,x_l)\mathcal T_{x_i}f(x_j,x_l),
\eeq
and $\mathcal L_u, \; \mathcal T_v$ are the usual left and right  translation operators, respectively.
\end{proposition}
{ From  the definition of the algebra product (\ref{qas}), the functions $f$ can be regarded as the algebra structure constants. Therefore, the form (\ref{function relation}) of the  universal  identity  (\ref{unidentity}) could be linked to the integrability condition of the differential equations associated with the considered generalized algebra.  However, at this stage of our study,  such an assertion deserves further  investigations.   }
%%%%%%%%%%%%%%%%%%%%%%%%%%%%%%%%%%%%%%%
%%%%%%%%%%%%%%%%%%%%%%%%%%%%%%%%%%%%%%%
\subsection{Phase space extension}
%%%%%%%%%%%%%%%%%%%%%%%%%%%%%%%%%%%%%%%
%%%%%%%%%%%%%%%%%%%%%%%%%%%%%%%%%%%%%%%
As known from \cite{kuper2}, the category of LSAs is closed with respect to the operation of phase-space extension, unlike the smaller category of associative algebras: if $\mathcal A$ is LSA then so is $T^\star \mathcal A= \mathcal A + \mathcal A^\star:$
\bea
\left(\begin{array}{c}e_{x_i}\\
e_{x_i}\rq{}
\end{array}\right)\star\rq{}\left(\begin{array}{c}e_{x_j}\\
e_{x_j}\rq{}
\end{array}\right)=\left(\begin{array}{c}e_{x_i}\star e_{x_j}\\
e_{x_i}\star e_{x_j}\rq{}-e_{x_j}\star e_{x_i}\rq{}
\end{array}
\right), \; e_{x_i}, e_{x_j}\in \mathcal A, \;e_{x_i}\rq{}, e_{x_j}\rq{}\in  \mathcal A^\star,
\eea
where, 
\beq
\langle e_{x_i}\star e_{x_j}\rq{}, e_{x_k} \rangle= -\langle e_{x_j}\rq{}, e_{x_i}\star e_{x_k}\rangle, \;  \; e_{x_i}, e_{x_k}\in \mathcal A, \; e_{x_j}\rq{}\in  \mathcal A^\star.
\eeq
The integrability of the hydrodynamical systems of the type \cite{kuper2}:
\beq\label{inth}
u_t= \rho u_{e_{x_i}} + u_{e_{x_i}}\star u
\eeq
 is preserved under such phase-space extensions. In (\ref{inth}), $\rho$ is the operator whose matrix elements are $\rho_{x_j}^{x_i}:$
\beq
\rho(e_{x_j})= \sum_{x_i}\rho_{x_j}^{x_i}e_{x_i},
\eeq
$\{e_{x_i}\}$ being a basis of $ \mathcal A,$ whose structure constants are $C_{x_jx_k}^{x_i}:$
\beq
e_{x_j}\star e_{x_k}=\sum_{x_i}C_{x_jx_k}^{x_i}e_{x_i}.
\eeq
Assume $\rho$ satisfies the following $\rho-$ compatibility equation (\ref{comp}), i. e.,
\bea
e_{x_i}\star \rho(e_{x_j}) - e_{x_j}\star \rho(e_{x_i})&=&   \rho\Big(e_{x_i}\star e_{x_j}-e_{x_j}\star e_{x_i}\Big)\cr
&:=&\rho\Big[f(x_i,x_j)e_{x_i+x_j}-f(x_j,x_i)e_{x_i+x_j}\Big]
\eea
By formulae (\ref{starnew1}),(\ref{starnew2}),(\ref{starnew3}), we realize a new left-symmetric multiplication in the following way:
\beq\label{starnew4}
e_{x_i}\star\rq{} e_{x_j}:=\Big(e_{x_i} +  \epsilon\rho(e_{x_i})\Big)\star e_{x_j}, \; \epsilon^2=0
\eeq
corresponding to the  phase-space extension 

\bea
\left(\begin{array}{c}e_{x_i}\\
e_{x_i}\rq{}
\end{array}\right)\star_1\rq{}\left(\begin{array}{c}e_{x_j}\\
e_{x_j}\rq{}
\end{array}\right)=\left(\begin{array}{c}e_{x_i}\star e_{x_j}\\
e_{x_i}\star e_{x_j}\rq{}-e_{x_j}\star e_{x_i}\rq{}
\end{array}
\right), \; e_{x_i}, e_{x_j}\in \mathcal A, \;e_{x_i}\rq{}, e_{x_j}\rq{}\in  \mathcal A^\star,
\eea
where

\bea
\langle e_{x_i}\star\rq{} e_{x_j}\rq{}, e_{x_k} \rangle&=& -\langle e_{x_j}\rq{}, e_{x_i}\star\rq{} e_{x_k}\rangle\cr
&=&-\langle e_{x_j}\rq{}, \Big(e_{x_i} +  \epsilon\rho(e_{x_i})\Big) \star e_{x_k}\rangle
\cr
&=&\langle \Big(e_{x_i} +  \epsilon\rho(e_{x_i})\Big)\star e_{x_j}\rq{},  \star e_{x_k}\rangle
\eea
implying
\beq
e_{x_i}\star\rq{} e_{x_j}\rq{}= \Big(e_{x_i} +  \epsilon\rho(e_{x_i})\Big)\star e_{x_j}\rq{}.
\eeq
Provided the former relation and taking into account the natural extension :
 $\mathcal A \rightarrow T^\star \mathcal A= \mathcal A + \mathcal A^\star: \rho \mapsto \rho_1$ should  satisfy:
%%%%%%%%%%%% Extended Virasoro algebra%%%%%%%%%%%%%
%%%%%%%%%%%%%%%%%%%%%%%%%%%%%%%%%%%%%%%
\bea
\left(\begin{array}{c}e_{x_i}\\
e_{x_i}\rq{}
\end{array}\right)\star_1\rq{}\left(\begin{array}{c}e_{x_j}\\
e_{x_j}\rq{}
\end{array}\right)=\left(\left(\begin{array}{c}e_{x_i}\\
e_{x_i}\rq{}
\end{array}\right)+\epsilon \rho_1\left(\begin{array}{c}e_{x_i}\\
e_{x_i}\rq{}
\end{array}\right)\right)\star_1\rq{}\left(\begin{array}{c}e_{x_j}\\
e_{x_j}\rq{}
\end{array}\right),
\eea
  we are in right to postulate the following  choice:
\begin{proposition}
%Let 
\beq\label{case1}
\rho_1: \left(\begin{array}{c}e_{x_i}\\
e_{x_i}\rq{}
\end{array}\right) \mapsto \left(\begin{array}{c}\rho (e_{x_i})\\
0
\end{array}\right)
\eeq
 satisfies the $\rho-$compatibility equation (\ref{comp}) in $T^\star \mathcal A.$
\end{proposition}
%is a good choice.
\proof
We have:

\bea\label{case2}
\rho_1\left(\left(\begin{array}{c}e_{x_i}\\
e_{x_i}\rq{}
\end{array}\right)\star_1\left(\begin{array}{c}e_{x_j}\\
e_{x_j}\rq{}
\end{array}\right) - 
\left(\begin{array}{c}e_{x_j}\\
e_{x_j}\rq{}
\end{array}\right)\star_1\left(\begin{array}{c}e_{x_i}\\
e_{x_i}\rq{}
\end{array}\right)
\right)=
\rho_1\left(\begin{array}{c}e_{x_i}\star e_{x_j}- e_{x_j}\star e_{x_i}\\
e_{x_i}\star e_{x_j}\rq{}- e_{x_j}\star e_{x_i}\rq{}
\end{array}\right)\nonumber\\=
\left(\begin{array}{c}\rho(e_{x_i}\star e_{x_j}- e_{x_j}\star e_{x_i})\\
0
\end{array}\right)
\eea

\bea\label{case3}
\left(\begin{array}{c}e_{x_i}\\
e_{x_i}\rq{}
\end{array}\right)\star_1\rho_1\left(\begin{array}{c}e_{x_j}\\
e_{x_j}\rq{}
\end{array}\right) - 
\left(\begin{array}{c}e_{x_j}\\
e_{x_j}\rq{}
\end{array}\right)\star_1\rho_1\left(\begin{array}{c}e_{x_i}\\
e_{x_i}\rq{}
\end{array}\right)
=
\left(\begin{array}{c}e_{x_i}\\
e_{x_i}\rq{}
\end{array}\right)\star_1\left(\begin{array}{c}\rho(e_{x_j})\\
0
\end{array}\right)\nonumber\\ - 
\left(\begin{array}{c}e_{x_j}\\
e_{x_j}\rq{}
\end{array}\right)\star_1\left(\begin{array}{c}\rho(e_{x_i})\\
0
\end{array}\right)= 
\left(\begin{array}{c}e_{x_i}\star \rho(e_{x_j})- e_{x_j}\star \rho(e_{x_i}),\\
0
\end{array}\right).
\eea
Therefore, the $\rho-$compatibility equation (\ref{comp}) also holds  in $T^\star \mathcal A.$
\qed

When $\rho: e_{x_i}\mapsto e_{x_i+x_0}$ for some fixed $x_0,$ the choice (\ref{case1}) reduces to a simpler relation:
\beq\label{case1'}
\rho_1: \left(\begin{array}{c}e_{x_i}\\
e_{x_i}\rq{}
\end{array}\right) \mapsto \left(\begin{array}{c} e_{x_i+x_0}\\
0
\end{array}\right).
\eeq
We infer the following result.

\begin{proposition}
Let $\rho: e_{x_i}\mapsto e_{x_i+x_0}.$ Then the relations (\ref{case2}) and (\ref{case3}) 
reduced to
\bea
%\left(\begin{array}{c} f(x_i, x_j)\Big[1-\mathcal E + \mathcal T_{x_0}(\mathcal E -1)\Big]\\
%0
%\end{array}\right)= 
\left(\begin{array}{c}\Big[ f(x_i, x_j) - f(x_j, x_i)\Big]e_{x_i+x_j+x_0}\\0
\end{array}\right) =  \left(\begin{array}{c}\Big[ f(x_i, x_j+x_0) - f(x_j,x_i+x_0)\Big]e_{x_i+x_j+x_0}\\0
\end{array}\right),
\eea
which is equivalent to
\bea
\left(\begin{array}{c} \Big[1-\mathcal E + \mathcal T_{x_0}(\mathcal E -1)\Big]f(x_i, x_j)\\
0
\end{array}\right)
= \left( \begin{array}{c}0\\0
\end{array}\right)
\eea
giving
\bea\label{case4}
 \mathcal T_{x_0}(1- \mathcal E)= (1- \mathcal E).
\eea
\end{proposition}

We thus recover the $\rho-$compatibility condition (\ref{case0}).

The phase-space extension of the system (\ref{inth}) reads:

\bea
\left(\begin{array}{c}u\\
u\rq{}
\end{array}\right)_t=  \rho_1\left(\begin{array}{c}u_{e_{x_i}}\\
u_{e_{x_i}}\rq{}
\end{array}\right) + \left(\begin{array}{c}u_{e_{x_i}}\\
u_{e_{x_i}}\rq{}
\end{array}\right)\star_1\left(\begin{array}{c}u\\
u\rq{}
\end{array}\right)=
\left(\begin{array}{c}\rho(u_{e_{x_i}})+  u_{e_{x_i}}\star u \\
u_{e_{x_i}}u\rq{}
\end{array}\right).
\eea

It is also worth noticing that, in the case of the left-symmetric double \cite{kuper2}: $\mathcal A^d= \mathcal A + \mathcal A,$ we get

\bea
\left(\begin{array}{c}e_{x_i}\\
e_{x_a}
\end{array}\right)\star_2\left(\begin{array}{c}e_{x_j}\\
e_{x_b}
\end{array}\right) =\left(\begin{array}{c}e_{x_i}\star e_{x_j} \\
e_{x_i}\star e_{x_b}-e_{x_j}\star e_{x_a}\end{array}\right)=\left(\begin{array}{c}f(x_i, x_j)e_{x_i+x_j}\\
f(x_i, x_b)e_{x_i + x_b}-f(x_j, x_a)e_{x_j + x_a}
\end{array}\right)
\eea
and if $\rho:\mathcal A \rightarrow \mathcal A$  is an operator of strong deformation, then so is $\rho_2: \mathcal A^d\rightarrow \mathcal A^d,$

\bea
\rho_2\left(\begin{array}{c}e_{x_i}\\
e_{x_a}
\end{array}\right)=\left(\begin{array}{c}\rho(e_{x_i}) + e_{x_g}e_{x_a}\\
e_{x_\mu}e_{x_i}+e_{x_\nu}e_{x_a}
\end{array}\right), \;\;e_{x_g}, e_{x_\mu}, e_{x_\nu}= \mbox{constants}
\eea
yielding 

\bea
\rho_2\left(\begin{array}{c}e_{x_i}\\
e_{x_a}
\end{array}\right)=\left(\begin{array}{c}e_{x_i+x_0} + e_{x_g}e_{x_a}\\
e_{x_\mu}e_{x_i}+e_{x_\nu}e_{x_a}
\end{array}\right), \;\; \mbox{when} \;\;\rho(e_{x_i}):= e_{x_i+x_0}.
\eea

Furthermore, as claimed in \cite{kuper2}, the formula (\ref{starnew4}) shows that, in addition to the integrable hydrodynamic hierarchy starting with the equation (\ref{inth}), i.e.,
\beq
u_t= \rho u_{e_{x_i}} + u_{e_{x_i}}\star u
\eeq
we have a second hierarchy, starting with the equation

\beq\label{inth2}
u_t= u_{e_{x_i}}\star\rq{}u=u_{e_{x_i}}\star u +\epsilon \rho (u_{e_{x_i}} )\star  u, \;\; \epsilon^2=0.
\eeq

%%%%%%%%%%%% Extended Virasoro algebra%%%%%%%%%%%%%
%%%%%%%%%%%%%%%%%%%%%%%%%%%%%%%%%%%%%%%
\section{ Centrally extended Virasoro algebra}
%%%%%%%%%%%%%%%%%%%%%%%%%%%%%%%%%%%%%%%
%%%%%%%%%%%%%%%%%%%%%%%%%%%%%%%%%%%%%%%
In this section we aim at investigating the quasi-associativity condition, the $3-$ary bracket and the fundamental identity
%, as well as the consequences of the hereditarity and $\rho-$ compatibilty equation
 for  the above generalized algebra in the particular case when $a=b=1$ and 

\beq
f({x_i}, {x_j})=  -{{ x_j(1+\epsilon x_j) }\over {1+\epsilon ( x_i+x_j)}}
+{1\over 2}\theta\Big [{x_i}^3 - {x_i} +(\epsilon - \epsilon^{-1}){x_i}^2\Big]\delta_{x_i+x_j} ^0,
\eeq

\bea\label{gvir0}
g({x_i}, {x_j})&=& ({x_i}-{x_j}) + \theta ({x_i}^3 - {x_i})\delta_{x_i +x_j}
%[e_{x_i}, \theta]&=& [ \theta, e_{x_i}]= 0
\eea

%\beq
%e_{x_i}\star e_{x_j}=  -{{ x_j(1+\epsilon x_j) }\over {1+\epsilon ( x_i+x_j)}}
%e_{x_i+x_j} +{1\over 2}\theta\Big [{x_i}^3 - {x_i} +(\epsilon - \epsilon^{-1}){x_i}^2\Big]\delta_{x_i+x_j} ^0
%\eeq
%and
%\beq
%e_{x_i}\star \theta=  \theta \star e_{x_i}=0.
%\eeq
This algebra corresponds to  the Virasoro algebra, also called central extension of the Witt algebra with the multiplication:
\bea\label{gvir0}
[e_{x_i}, e_{x_j}]&=&g ({x_i}-{x_j}) e_{x_i +x_j},\cr
e_{x_i}\star \theta&=& \theta\star e_{x_i}= 0
\eea
coming from the commutator
\bea\label{com0}[e_{x_i}, e_{x_j}]&= & e_{x_i}\star e_{x_j}-  e_{x_j}\star e_{x_i}\cr
e_{x_i}\star e_{x_j}&=&f({x_i}, {x_j})e_{x_i+x_j}
\eea
where $  ({x_i}, {x_j})\in \mathbb Z^{2}.$
%The Virasoro algebra is widely used in conformal field theory and in string theory. 

\begin{proposition}For the centrally extended Virasoro algebra,
\begin{itemize}
\item[(i)] 
The skew-symmetry condition (\ref{sk}) 
is equivalent to the system of equations
\bea
\begin{cases}  x_i+x_j +\epsilon\Big( x_i^2+x_j^2\Big)=0\\
x_i^3+x_j^3- (x_i+x_j) + \Big(\epsilon -\epsilon^{-1}\Big)\Big(x_i^2+x_j^2\Big)= 0;
\end{cases}
\eea

\item[(ii)] The Jacobi identity
\beq
\Big[ [e_{x_i}, e_{x_j}],e_{x_k}\Big] + \Big[ [e_{x_j}, e_{x_k}],e_{x_i}\Big] + \Big[ [e_{x_k}, e_{x_i}],e_{x_j}\Big]= 0
\eeq
is identified to the condition
\beq\label{J1}
{\mathbb J}_{ij}^k + {\mathbb J}_{jk}^i + {\mathbb J}_{ki}^j=0
\eeq
where
\bea
{\mathbb J}_{ij}^k:=\Big[({x_i}-{x_j})&+& \theta ({x_i}^3 - {x_i})\delta_{x_i +x_j}\Big]
\Big[({x_i+x_j}-{x_k}) \cr & +& \theta\Big (({x_i}+x_j)^3 - ({x_i}+x_j)\Big)\delta_{x_i +x_j+x_k}\Big];
\eea

\item[(iii)] The derivation property, i.e.,

\beq
[e_{x_i}, e_{x_j}\star e_{x_k}]: = e_{x_j}\star [e_{x_i},  e_{x_k}] +[e_{x_i}, e_{x_j}]\star e_{x_k}
\eeq
leads to

\bea
&& -{{ x_k(1+\epsilon x_k) }\over {1+\epsilon ( x_j+x_k)}}
  ({x_i}-({x_j}+x_k)) e_{x_i +x_j+x_k}
 + \theta ({x_i}^3 - {x_i})\delta_{x_i +x_j+x_k}^0  ={1\over {1+\epsilon ( x_i+x_j+x_k)}}\cr
&\times&
 \Big[ - (x_i^2-x_k^2)- x_k(x_i-x_j)-\epsilon [(x_i^2-x_k^2)(x_i+x_k)+x_k^2(x_i-x_j)]\Big]e_{x_i+x_j+x_k}\cr
& +&{1\over 2}\theta\Big [({x_j}^3 - {x_j} )(x_i-x_k) + (x_i-x_j)[(x_i+x_j)^3-(x_i+x_j)]\cr
&+&(\epsilon - \epsilon^{-1})[{x_j}^2(x_i-x_k) +(x_i-x_j)(x_i+x_j)^2]\Big]\delta_{x_i+x_j+x_k} ^0.
%%\Big]
%%\cr
%%&+&\Big[({x_i}-{x_j}) e_{x_i +x_j} + \theta ({x_i}^3 - {x_i})\delta_{x_i +x_j}^0\Big]
%%\Big[ -{{ x_k(1+\epsilon x_k) }\over {1+\epsilon ( x_i+x_j+x_k)}}
%%e_{x_i+x_j+x_k}\cr
%%& +&{1\over 2}\theta\Big [{(x_i}+x_j^3 - {(x_i+x_j)} +(\epsilon - \epsilon^{-1})(x_i-x_j)%(x_i+x_j)^2\Big]\delta_{x_i+x_j+x_k} ^0\Big].
\eea

\end{itemize}
%Exploiting the relation 
%%(\ref{fg})
% between the functions $f$ and $g,$ i.e.,
%
%\beq\label{fg}
%g(x_i, x_j)= af(x_i, x_j) - bf({x_j, x_i}).
%\eeq
%
%we can re-express the criterion for the Jacobi identy by the following result:
%\bea\label{J2}
%a^2\Big({\mathbb T}_{ij}^k + {\mathbb T}_{jk}^i + {\mathbb T}_{ki}^j\Big) + b^2 \Big({\mathbb G}_{ji}^k + {\mathbb G}_{ik}^j + {\mathbb G}_{kj}^i\Big)
%-ab\Big({\mathbb G}_{ij}^k + {\mathbb G}_{jk}^i + {\mathbb G}_{ki}^j 
%+{\mathbb T}_{kj}^i + {\mathbb T}_{ji}^k + {\mathbb T}_{ik}^j\Big)=0
%\eea
%where
%\beq
%{\mathbb T}_{ij}^k := f({x_i}, {x_j}) f({x_i + x_j}, {x_k}),\;\;\;\;{\mathbb G}_{ij}^k =:f({x_i}, {x_j}) f({x_k,  x_i+x_j}).
%\eeq
%
%
%
%

\end{proposition}
\subsection{Quasi-associativity condition}

We answer the question: Does it exist a necessary and sufficient condition for this algebra to be
 a quasi-associative algebra with the multiplication
%\bea\label{qas0}
%% e_{x_i}\star e_{x_j}&=&f({x_i,x_j})e_{x_i+x_j}\cr
%%e_{x_i}\star \theta&=&  \theta \star  e_{x_i}= 0?
%\eea
\bea\label{qas0}
e_{x_i}\star e_{x_j}&=&  -{{ x_j(1+\epsilon x_j) }\over {1+\epsilon ( x_i+x_j)}}
e_{x_i+x_j} +{1\over 2}\theta\Big [{x_i}^3 - {x_i} +(\epsilon - \epsilon^{-1}){x_i}^2\Big]\delta_{x_i+x_j} ^0\cr
%\eeq
%and
%\beq
e_{x_i}\star \theta&=&  \theta \star e_{x_i}=0?
\eea
\begin{theorem}
The algebra defined with the multiplication rule (\ref{qas0}) is neither associative, nor left-symmetric.
\end{theorem}

\proof
By direct computation, we find:

\bea
&&e_{x_i}\star (e_{x_j}\star e_{x_k})-(e_{x_i}\star e_{x_j})\star e_{x_k}\cr
&=&{{ x_k(1+\epsilon x_k) }\over {1+\epsilon ( x_i+x_j+x_k)}}
\Big( {{x_k+  \epsilon\Big[(x_i+x_j)(x_j+x_k)- x_j^2\Big]}\over{1+\epsilon ( x_i+x_j)}}\Big)
e_{x_i+x_j+x_k}\cr
&-&{\theta \over {2}}\Big({{ x_k(1+\epsilon x_k) }\over {1+\epsilon ( x_j+x_k)}}\Big[{x_i}^3 - {x_i} +(\epsilon - \epsilon^{-1}){x_i}^2\Big]\cr
&-&  {{ x_j(1+\epsilon x_j) }\over {1+\epsilon ( x_i+x_j)}} \Big[({x_i+x_j})^3 - ({x_i+x_j}) +(\epsilon - \epsilon^{-1})({x_i+x_j})^2\Big]\Big)\delta_{x_i+x_j+x_k}^0
\eea

while

\bea
&&e_{x_j}\star (e_{x_i}\star e_{x_k})-(e_{x_j}\star e_{x_i})\star e_{x_k}\cr
&=&{{ x_k(1+\epsilon x_k) }\over {1+\epsilon ( x_i+x_j+x_k)}}
\Big( {{x_k+  \epsilon\Big[(x_i+x_k)(x_i+x_j)- x_i^2\Big]}\over{1+\epsilon ( x_i+x_j)}}\Big)
e_{x_i+x_j+x_k}\cr
&-&{\theta \over {2}}\Big({{ x_k(1+\epsilon x_k) }\over {1+\epsilon ( x_i+x_k)}}\Big[{x_j}^3 - {x_j} +(\epsilon - \epsilon^{-1}){x_j}^2\Big]\cr
&-&  {{ x_i(1+\epsilon x_i) }\over {1+\epsilon ( x_i+x_j)}} \Big[({x_i+x_j})^3 - ({x_i+x_j}) +(\epsilon - \epsilon^{-1})({x_i+x_j})^2\Big]\Big)\delta_{x_i+x_j+x_k}^0\cr
&\ne& e_{x_i}\star (e_{x_j}\star e_{x_k})-(e_{x_i}\star e_{x_j})\star e_{x_k}. 
\eea
\qed

It is worth noticing that for $i=j,$ this algebra becomes a left-alternative algebra as required by the general formalism developed in the previous section.

\subsection{$3-$ary bracket and fundamental identity}
The $3-$ary bracket, defined by the relation (\ref{3brac}), i.e.,
\beq
[e_{x_i}, e_{x_j},e_{x_k}]:= e_{x_i}\star [e_{x_j},e_{x_k}] + e_{x_j}\star [e_{x_k},e_{x_i}] + e_{x_k}\star [e_{x_i},e_{x_j}]\nonumber
\eeq
leads to the expression
\bea
[e_{x_i}, e_{x_j},e_{x_k}]&=&- {1\over {1+\epsilon ( x_i+x_j+x_k)}} \epsilon\Big[(x_j^2 -x_i^2)x_k +(x_i^2 -x_k^2)x_j +(x_k^2 - x_j^2)x_i\Big] e_{x_i+x_j+x_k} 
\cr 
&+&
{{\theta} \over {2}}
\Big[({x_i}^3+ {x_j}^3+{x_k}^3) (1+\epsilon - \epsilon^{-1})- ({x_i} +{x_j} +{x_k})\Big]
\delta_{x_i+x_j+x_k}^0.
\eea

Defining the fundamental identity, (also called Filippov identity), in this case as:
\bea
%&&
\Big[{x_i}, e_{x_j}, [e_{x_k}, e_{x_l},e_{x_m}]\Big]&:= &\Big[[{x_i}, e_{x_j}, e_{x_k}], e_{x_l},e_{x_m}\Big]
%\cr 
%&+&
+\Big[{x_k},[ e_{x_i}, e_{x_j}, e_{x_l}],e_{x_m}\Big]\cr
&+&\Big[{x_k}, e_{x_l}, [e_{x_i}, e_{x_j},e_{x_m}]\Big],
\eea 
implies cumbersome functional equations:
\bea
&&{{ \Big[(x_l^2 -x_k^2)x_m +(x_k^2 -x_m^2)x_l +(x_m^2 - x_l^2)x_k\Big] }\over {1+\epsilon ( x_k+x_l+x_m)}}
\Big[(x_j^2 -x_i^2)(x_k +x_l+x_m) 
\cr
&+&
\Big (x_i^2 -(x_k+x_l+x_m)^2\Big)x_j
+\Big((x_k+x_l+x_m)^2 - x_j^2\Big)x_i\Big]\cr
&=&
{{\Big [(x_j^2 -x_i^2)x_k +(x_i^2 -x_k^2)x_j +(x_k^2 - x_j^2)x_i\Big] }\over {1+\epsilon ( x_i+x_j+x_k)}}
%&\times&
\Big[(x_l^2 -(x_i+x_j+x_k)^2)x_m \cr
&+& \Big((x_i+x_j+x_k)^2-x_m^2\Big)x_l
%\cr
%&+&
+ (x_m^2 -x_l^2)(x_i+x_j+x_k)\Big]
\cr
&+&
{{\Big [(x_j^2 -x_i^2)x_l +(x_i^2 -x_l^2)x_j +(x_l^2 - x_j^2)x_i\Big] }\over {1+\epsilon ( x_i+x_j+x_l)}}
\Big[\Big((x_i+x_j+x_l)^2 -x_k^2\Big)x_m\cr
&+&(x_k^2-x_m^2)(x_i+x_j+x_l)
+(x_m^2-(x_i+x_j+x_l)^2)x_k\Big]
 \cr
&+&
{{\Big [(x_j^2 -x_i^2)x_m +(x_i^2 -x_m^2)x_j +(x_m^2 - x_j^2)x_i\Big] }\over {1+\epsilon ( x_i+x_j+x_m)}}
\Big[(x_l^2-x_k^2)(x_i+x_j+x_m)\cr
&+&(x_k^2-(x_i+x_j+x_m)^2) x_l
%\cr
+\Big((x_j +x_j+x_m)^2-x_l^2\Big)x_k\Big]
%
%&+& ((x_i+x_j+x_k)^2-x_m^2)x_l
%%\cr
%%&+&
%+ (x_m^2 -x_l^2)(x_i+x_j+x_k)]
%+((x_k+x_l+x_m)^2 - x_j^2)x_i]
\eea
and
\bea\label{epsilon }
&&\Big[(x_i+x_j+x_k)^3+(x_j+x_k+x_l)^3 +(x_i+x_j+x_m)^3 -(x_k+x_l+x_m)^3\cr&+& 2(x_k^3+x_l^3+x_m^3)
- (x_i^3+x_j^3)\Big]
%\cr
%&\times&
(1+\epsilon+\epsilon^{-1})-2(x_i+x_j+x_k+x_l+x_m)=0.
\eea
These functional equations reduce to the simpler relation (\ref{epsilon }) when $\epsilon^2= 0.$
%\bea
%&&\left(x_i^3+x_j^3+(x_k+x_l+x_m)^3\right)(1+\epsilon+\epsilon^{-1})+2(x_i+x_j+x_k+x_l+x_m)\cr
%&=&\left[(x_i+x_j+x_k)^3+(x_j+x_k+x_l)^3 +(x_i+x_j+x_m)^3 + 2(x_k^3+x_l^3+x_m^3)\right]\cr
%&\times&(1+\epsilon+\epsilon^{-1})
%\eea
%%%%%%%%%%%%%%%%%%%%%%%%%%%%%%%%%%%%%%%%%%%%
%\end{document}
%%%%%%%%%%%%%%%%%%%%%%%%%%%%%%%%%%%%%%%%%%%%
\section{$\mathbb L_k-$ infinite dimensional Lie algebra of polynomial vector fields on the real line $\mathbb R^1$}
Consider the algebra $\mathbb L_k-$  as the infinite dimensional Lie algebra of polynomial vector fields on the real line $\mathbb R^1.$ 
%with a zero in $x=0$ of order not less than $k+1.$
Let us define this algebra by the infinite basis $\{e_i\}:$
\beq
e_i=x^{i+1}{d\over{dx}},\;\; i\in \mathbb N
\eeq
with the commutator
\beq
[e_i,e_j]= e_i\star e_j- e_j\star e_i= (j-i)e_{i+j}
\eeq
with the multiplication
\beq\label{product}
e_p\star e_q= (q+1)e_{p+q} + e_{p+q+1}{d\over{dx}}.
\eeq
Here the $\star-$ multiplication is nothing but the ordinary operators product.

One can easily prove by direct computation that this algebra endowed with the product (\ref{product}) is an associative algebra, i.e., its associator is equal to zero. Indeed,
\bea
e_p\star(e_q\star e_r)&=&(e_p\star e_q)\star e_r)\cr
&=&(r+1)(q+r+1)e_{p+q+r} +(q+2r+3)e_{p+q+r+1}{d\over{dx}}+e_{p+q+r+2}{d^2\over{dx^2}}.
\eea
Further the corresponding Nambu brackets are null, i.e.,
\beq
[e_{p}, e_{q},e_{r}]:= e_{p}\star [e_{q},e_{r}] + e_{q}\star [e_{r},e_{p}] + e_{r}\star [e_{p},e_{q}]=0,
\eeq
that is the Jacobi identity is automatically satisfied. Thus we have a null $3-$ algebra for an infinite set of non-trivial noncommuting oscillator charges. The Filippov condition is trivially satisfied in this case, i.e.,
\beq
\Big[e_{p}, e_{q},[e_{r},e_{s},e_{t}]\Big]:= \Big[[e_{p}, e_{q},e_{r}],e_{s},e_{t}\Big]+\Big [e_{r}, [e_{p},e_{q},e_{s}],e_{t}\Big]+\Big[e_{r}, e_{s},[e_{p},e_{q},e_{t}]\Big].
\eeq
The Bremner operator, also called the associative operator, of course,   perfectly works, i.e.,
\beq
\Big[\Big[e_{p}, [e_{q}, e_{r},e_{s}],e_{t}\Big], e_u, e_v\Big]:=\Big [[e_{p}, e_{q},e_{r}],[e_{s},e_{t},e_u], e_{v}\Big]
\eeq
as a consequence of the associativity. Moreover, the skew-symmetry property is
obeyed by the definition of the product, i.e.,
\beq
[e_i,e_j]= e_i \star e_j- e_j\star e_i=-[e_j,e_i]= - (j-i)e_{i+j}.
\eeq
Provided  the Jacobi identity and skew-symmetry properties are satisfied, this algebra is made into a Lie algebra structure.
On the other hand,

\beq\label{rhs}
[e_{p}, e_{q}\star e_{r}]= \Big[ (r+1)(q+r+1)-(p+1)(p+r+1)\Big]e_{p+q+r}+(q+r-2p)e_{p+q+r+1}{d\over{dx}},
\eeq

\beq\label{lhs1}
e_{q}\star[e_{p},  e_{r}]= (p+r+1)(r-p)e_{p+q+r}+(r-p)e_{p+q+r+1}{d\over{dx}},
\eeq

\beq\label{lhs2}
[e_{p},  e_{q}]\star e_q= (r+1)(q-p)e_{p+q+r}+(q-p)e_{p+q+r+1}{d\over{dx}}.
\eeq
Summing the relations (\ref{lhs1}) and (\ref{lhs2}), we find that
 
\beq
[e_{p}, e_{q}\star e_{r}]= e_{q}\star[e_{p},  e_{r}]+[e_{p},  e_{q}]\star e_q,
\eeq

showing that the derivation property is satisfied, what makes $\left(\mathbb L_k, [.,]\right)$ into a Poisson structure.

These properties induce the following consequences:

\begin{proposition}
Let $L_{e_q}$ and $R_{e_q}$ be the left and right multiplication operators by $e_q$,  (for some fixed  $e_q\in \mathbb L_k),$  defined, respectively, as:

\beq
L_{e_q}(e_p)= e_q\star e_p, \; \; R_{e_q}(e_p)=  e_p\star e_q, \forall e_p\in \mathbb L_k.
\eeq
Then the following relations hold, for all $e_p, e_q \in  \mathbb L_k:$
\begin{itemize}
\item
\beq
[L_{e_p}, L_{e_q}]=L_{[e_p,e_q]}
\eeq
\item
\beq
R_{[{e_p}, {e_q}]}(.)=(.) \star[e_p,e_q]= -(.)\star[R_{e_p}, R_{e_q}]
\eeq

\item
\beq
{[L_{e_p}, R_{e_q}]}(.)=e_p \star[(.),e_q]=0
\eeq

\item
\beq
{[R_{e_p}, {e_q}]}(.)=e_q \star[e_p, (.)]
\eeq

\item
\beq
[R_{e_p}R_{e_q}+R_{e_p\star e_q}](.)=(.)\star [R_{e_p}(e_q) + R_{e_q}(e_p)].
\eeq
\end{itemize}
\end{proposition}
Let us mention an interesting identity of general interest:
\beq
\Big[e_{p}, [e_{q},e_{r}]\Big]+\Big[e_{r}, [e_{p},e_{q}]\Big] =e_q\star [e_{p},e_{r}]-[e_{p},e_{r}]\star e_q,
\eeq
valid whatever the $\star-$product, with the usual commutator.

\section{Concluding remarks}
{ In this paper, we have discussed the appearence of  left symmetric algebras  in a generalized Virasoro algebra.
We have provided the necessary and sufficient condition for this algebra to be
 a quasiassociative algebra. 
The criteria of skew-symmetry, derivation and Jacobi identity making this algebra  a Lie  algebra   have been derived. Coboundary operators are defined, the $2-$cocycle and coboundary are discussed. We have deduced the hereditary operator and its generalization to the corresponding $3-$ary bracket.  Further, we have derived the so-called $\rho-$compatibility equation, and performed a  phase-space extension. 
Concrete relevant particular cases  have also  been investigated and discussed. 

This study brings some interesting questions to light which merit a separate in-depth   treatment.  For instance, new examples of nonlinear systems associated with the considered generalization of the  Virasoro algebra  may exist, but their full investigation remains totally open. Besides, a detailed analysis of the main properties of $3-$algebras on the basis of  definition (\ref{3bracket0}) or (\ref{3bracket00}) is also of great importance.  These topics will  be  the core of our  forthcoming works}. 
\begin{center}
{\bf Acknowledgements}
\end{center}

 The authors are grateful to anonymous referees for their useful comments which permit to 
 improve the paper.  The work of  MNH is partially supported by the Abdus Salam International Centre for Theoretical Physics (ICTP, Trieste, Italy) through the Office of External Activities (OEA)-Prj-15. The ICMPA is also in partnership with the Daniel Iagolnitzer Foundation (DIF), France.
MNH and PG acknowledge
 the staff of  the Centre Interfacultaire Bernoulli for the hospitality during their stay from October to December 2014 at the  Ecole Polytechnique F\'ed\'erale de Lausanne, Switzerland. 
MNH thanks Professor Valentin Ovsienko for  his useful comments and suggestions.

\vspace{0.5cm}

\end{document}